# Fuzzy Lifetime Analysis of a Fault-Tolerant Two-Phase Interleaved Converter

Tohid Rahimi, Masoud Farhadi, *IEEE Student Member*, Ka Hong Loo, *IEEE Senior Member*, Josep Pou, *IEEE Fellow*

*Abstract* -- Interleaved converters are used in photovoltaic (PV) applications to handle high power conditions with high reliability. To improve the reliability of these converters, redundant switch configuration can be employed which reduce the failure rates of the power switches significantly. However, evaluation reliability of the interleaved converters equipped with redundant switch configurations may be complex. This paper aims to simplify the reliability Analysis of interleaved converters considering mission profile and redundant switch configuration. Different possible configurations of the studied converter are shown as Markov chain states in the proposed method, which simplify the reliability and failure rates Analysis. The effect of different parameters such as the converter power level, switch configuration type, and operation modes are derived and a better insight into the effectiveness of the switch configuration on improving the reliability is provided.

*Index Terms*-- Interleaved converters, photovoltaic applications, redundant switch configurations, reliability.

## I. INTRODUCTION

POWER electronic converters are used in wide ranges of power levels, voltage, and current rates for distributed energy applications. With the progress in manufacturing technology of power converters, grid-connected interfaces for the distributed power sources have been developed in recent decades [1]-[5]. By increasing roles of power converters in vital and high-power applications, reliability of these systems is becoming an important issue [6]-[9]. The semiconductor power switches that are the most vulnerable elements must be investigated from the reliability point of view [10]-[14]. One of the solutions is employing redundant switch configuration. However, the reliability Analysis of converters equipped with redundant switch configurations are complex. In [15], the mean time to failure (MTTF) of two- and three-phase interleaved boost converters is evaluated to determine the best configuration with considering the cost and reliability. In [16], the fault-tolerant operation of a single-switch converter using redundant switches is realized. However, the reliability Analysis of converters equipped with redundant switch configurations may be complex.

A well-known method for evaluating the lifetime of converters is the Markov-based analysis. In this method, the reliability and lifetime of a system are analyzed in the useful period of that system [17], [18]. Thus, the failure rates for special power and ambient conditions can be considered constant. There are different uncertainties in the manufacturing process of the devices which must be considered in the lifetime evaluation [19]. Fuzzy logic is a power full tool that can be used in reliability and lifetime evaluation in many industry applications. Fuzzy-based reliability evaluation of transmission system using fuzzy Markov model [20], a substation automation case study [21], [22], hybrid molten carbonate fuel cell (MCFC) and gas turbine system [23], are addressed in previous papers. An extended fuzzy-based fault tree analysis [24], reliability analysis of a robotic system using fuzzy numbers [25], and the fuzzy probability of unit reliability [26] are investigated in previous papers. In [27], to model uncertainty in load forecast, a fuzzy membership function of peak load is proposed. A fuzzy-Markov model is applied in [28] to incorporate fuzzy mean time to failure and fuzzy mean time to repair. However, fuzzy reliability analysis of power electronic converters using Markov chain theory is rarely discussed in previous papers. In this paper, the Markov model is based on the two-phase converter's electric model. This model helps the researchers understand how to handle the faulty switches. The converter is equipped with redundant power switch configurations. The fuzzy-logic-based Markov model is a powerful analysis tool that covers the reliability Analysis gap that is not considered in other literature.

## II. CASE STUDY WITH REDUNDANT SWITCH CONFIGURATION

The conventional two-phase dc-dc boost converter is considered in this paper, as shown in Fig. 1(a).

Interleaved operation has attractive benefits on efficiency, size, thermal management, and ripple cancellation input current. Fault detection and isolation mechanism for each power switch can be implemented by using the auxiliary switches ($S_{1a}$, $S_{2a}$). These auxiliary switches can manage the faulty conditions appropriately. Parallel and standby switch configurations are shown in Fig. 1(b). It is assumed that the converter can operate in one-phase operation mode.

---
Tohid Rahimi is with School of Electrical Engineering Shandong University Jinan, China (email: rahimitohid@yahoo.com).

Masoud Farhadi is with the Department of Electrical and Computer Engineering, The University of Texas at Dallas, TX, 75080 USA (e-mail: masoud.farhadi@utdallas.edu).

Ka Hong Loo Department of Electronic and Information Engineering The Hong Kong Polytechnic University, Hong Kong (email: kh.loo@polyu.edu.hk).

Josep Pou is with School of Electrical and Electronic Engineering Nanyang Technology University Singapore, Singapore (email: josep.pou@ieee.org).

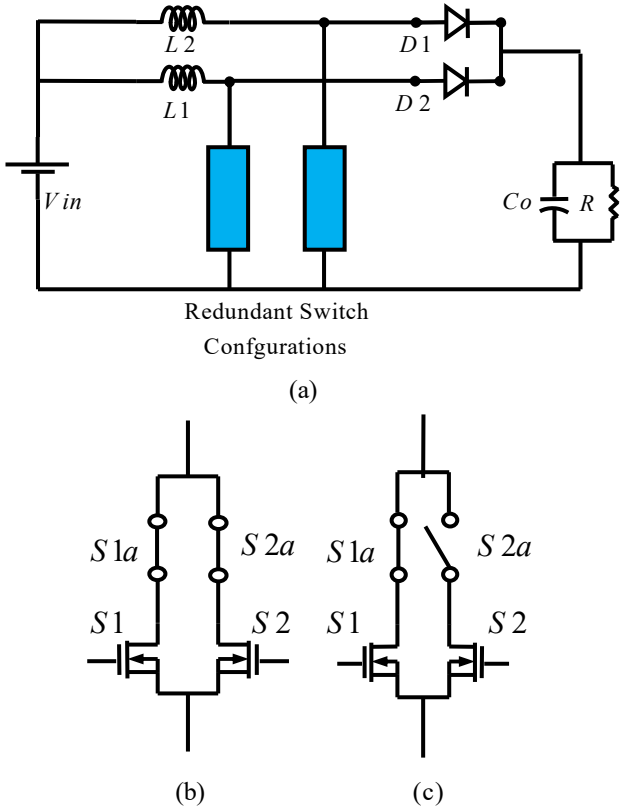

Fig. 1. (a) two-phase interleaved boost converter and redundant switch configurations. (b) Parallel and (c) standby configurations.

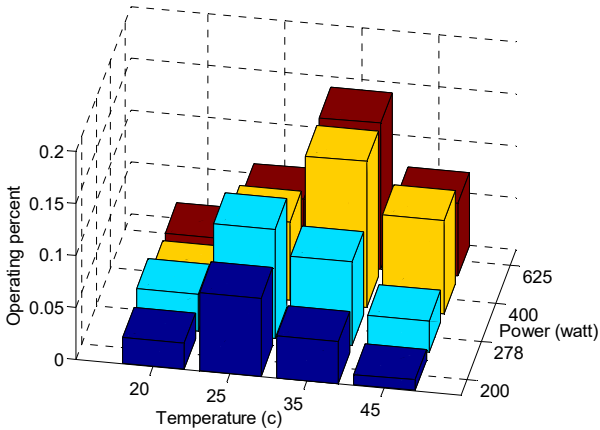

Fig. 2. Ambient temperature and output PV power distributions.

However, the junction temperatures of the switches should be calculated for this operation mode to ensure that the power switches operate in safe conditions.

Failure rates of power switches depend on the power dissipation and ambient temperature. Fig. 2 shows the clustered converter power level and temperature. This 3D-figure shows different monitored ambient temperature and power ranges and the corresponding percentage values of all conditions during a day. In fact, this temperature and power distribution profile is used to accurately estimate thermal stresses on the power switches. Considering participation of each power and temperature state in the failure rate calculations, the following equation can be obtained:

$$\lambda = \sum_{i=1}^{n} \lambda_i \times \mu_i \qquad (1)$$

where $\mu_i$ is the probability and $\lambda_i$ is the failure rate of the power switch in the state (i), respectively.

## III. MTTF ANALYSIS

In this paper, a two-phase interleaved boost converter is evaluated from the point of view of MTTF. Figs. 3 and 4 show the Markov reliability modes of the case study equipped with parallel and standby switch configurations. As illustrated in these figures, three possible operation states can be defined for the case study as follows:

- Three phases are healthy.
- One phase has failed. Then, the faulty phase is cleared, and the system operates as a two-phase converter.
- All phases failed.

The failure rate of the system when operating as a *k*-phase converter, is $\lambda^{(k)}$ and it is constant all the time. In the case of a fault occurrence, the probability of the mechanism that manages the fault is considered by $P_C$. For example, according to Fig. 4, the rate of transition from state (two-phase converter) to state (one-phase converter) is $2P_c \lambda^{(2)}$ in the Markov chain. If the strategy of the fault management fails, the probability of the transition from state (two-phase converter) to (fail state) is $2(1-P_c)\lambda^{(2)}$.

The base failure rate is considered as a fuzzy function as shown in Table I. With respect to Table I, the ambient conditions and PV power profile (Fig. 2), the failure rates of power switches in the used redundant configuration can be obtained. As seen in Fig. 3 and 4, to calculate the MTTF values of the two-phase converter, $\lambda_{H(i)}^{(2)}$, $\lambda_{F(i)}^{(2)}$, $\lambda_{H(i)}^{(1)}$, and $\lambda_{F(i)}^{(1)}$ must be determined for each power and temperature state. These parameters are fuzzy and can be extracted as

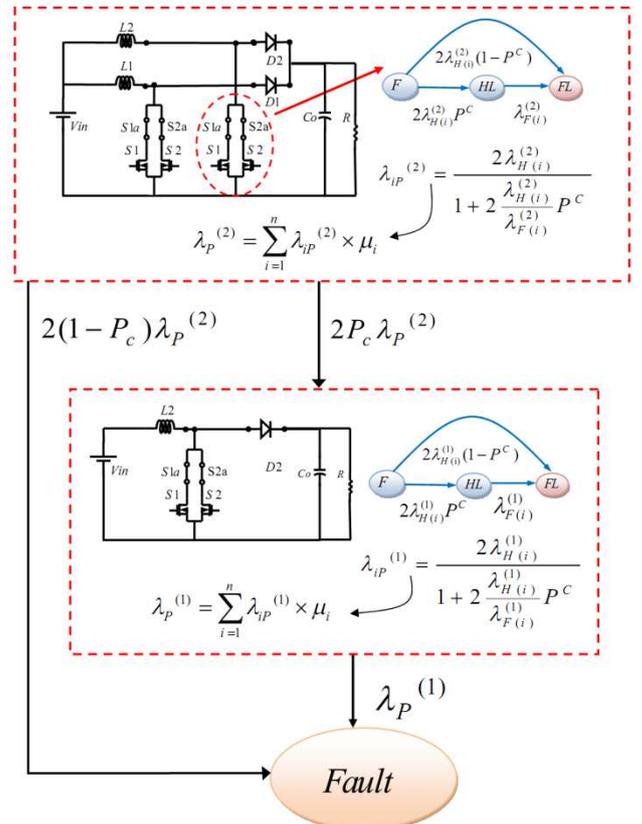

Fig. 3. Markov chain and reliability curves of a two-phase converter with parallel switch structure.

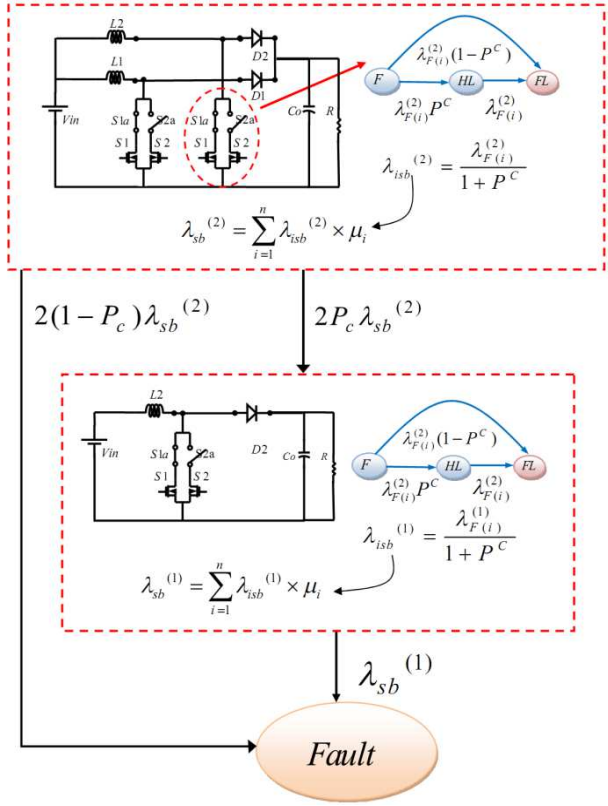

Fig. 4. Markov chain and reliability curves of a two-phase converter with standby switch structure.

follow:

$$\tilde{\lambda}_{H(i)}^{(k)} = \pi_Q \pi_A \pi_E \times \pi_{TH(i)}^{(k)} \times \tilde{\lambda}_b \quad (2)$$

$$\tilde{\lambda}_{F(i)}^{(k)} = \pi_Q \pi_A \pi_E \times \pi_{TF(i)}^{(k)} \times \tilde{\lambda}_b \quad (3)$$

For each state, the probabilities of parallel; and standby fails are calculated as follow:

$$\tilde{\lambda}_{iP}^{(k)} = \frac{2\tilde{\lambda}_{H(i)}^{(k)}}{1+2\frac{\tilde{\lambda}_{H(i)}^{(k)}}{\tilde{\lambda}_{F(i)}^{(k)}}\tilde{P}^C} = (ap_{\lambda_i^{(k)}}, bp_{\lambda_i^{(k)}}, cp_{\lambda_i^{(k)}}) \quad (4)$$

$$\tilde{\lambda}_{isb}^{(k)} = \frac{\tilde{\lambda}_{F(i)}^{(k)}}{1+\tilde{P}^C} = (as_{\lambda_i^{(k)}}, bs_{\lambda_i^{(k)}}, cs_{\lambda_i^{(k)}}) \quad (5)$$

Hence, the total failure rates of the converter in each operation mode for parallel and standby-based converter can be written as follow:

$$\tilde{\lambda}_P^{(2)} = \sum_{i=1}^n \tilde{\lambda}_{iP}^{(2)} \times \mu_i \quad (6)$$

$$\tilde{\lambda}_P^{(1)} = \sum_{i=1}^n \tilde{\lambda}_{iP}^{(1)} \times \mu_i \quad (7)$$

$$\tilde{\lambda}_{sb}^{(2)} = \sum_{i=1}^n \tilde{\lambda}_{isb}^{(2)} \times \mu_i \quad (8)$$

$$\tilde{\lambda}_{sb}^{(1)} = \sum_{i=1}^n \tilde{\lambda}_{isb}^{(1)} \times \mu_i \quad (9)$$

To calculate the MTTF of the two-phase converter, the probability of successful operation of fault manager is defined as seen in Table I. Therefore:

$$\widetilde{MTTF}_P^{(2)} = \frac{1 + \frac{2\tilde{\lambda}_P^{(2)}}{\tilde{\lambda}_P^{(1)}}\tilde{P}_c}{2\tilde{\lambda}_P^{(2)}} \quad (10)$$

$$\widetilde{MTTF}_{sb}^{(2)} = \frac{1 + \frac{2\tilde{\lambda}_{sb}^{(2)}}{\tilde{\lambda}_{sb}^{(1)}}\tilde{P}_c}{2\tilde{\lambda}_{sb}^{(2)}} \quad (11)$$

Table I. Fuzzy membership

| Base failure | Probability of successful operation of fault manager |
|---|---|
| $\tilde{\lambda}_b$, triangular membership with vertices at 0.01, 0.012 (peak, 1/year), 0.014 | $\tilde{P}_C$, triangular membership with vertices at 0.7, 0.8 (peak), 0.9 per-unit |

## IV. RESULTS

In each operation model of the converter, $\tilde{\lambda}_{iP}^{(k)}$ and $\tilde{\lambda}_{isb}^{(k)}$ are triangular memberships that depend on three parameters as given by (4) and (5). Figs. (5) and (6) show the ($ap_{\lambda_i^{(k)}}, bp_{\lambda_i^{(k)}}, cp_{\lambda_i^{(k)}}$ for parallel-based converter) and ($as_{\lambda_i^{(k)}}, bs_{\lambda_i^{(k)}}, cs_{\lambda_i^{(k)}}$ for standby-based converter) values for each power and temperature states when the converter operates in one phase mode operation. Figs. (7) and (8) show $ap_{\lambda_i^{(k)}}, bp_{\lambda_i^{(k)}}, cp_{\lambda_i^{(k)}}$ for parallel-based converter, and $as_{\lambda_i^{(k)}}, bs_{\lambda_i^{(k)}}, cs_{\lambda_i^{(k)}}$ for standby-based converter values for each power and temperature states when the converter operates in two-phase operation mode.

With respect to Figs. 5-8, and considering (6)-(9), the fuzzy membership of the total failure rates of power switches in each operation mode can be calculated as follows:

$$\tilde{\lambda}_P^{(2)} = (1.0335 \quad 1.8564 \quad 3.2574) \quad (12)$$

$$\tilde{\lambda}_P^{(1)} = (1.6696 \quad 2.9605 \quad 5.1197) \quad (13)$$

$$\tilde{\lambda}_{sb}^{(2)} = (1.1793 \quad 1.4937 \quad 1.8452) \quad (14)$$

$$\tilde{\lambda}_{sb}^{(1)} = (2.3718 \quad 3.0043 \quad 3.7112) \quad (15)$$

The fuzzy memberships of different parameters that described in this paper are used to extract fuzzy curves of MTTF function of the studied topologies are extracted, as seen in Fig. 9. These fuzzy MTTF curves show MTTF distribution of the two-phase converter with the referred redundant power switch configurations. Defuzzification is the final step in Fuzzy-based reliability evaluation. Different defuzzification methods can be implemented. In this paper, the method proposed in [29] is used to map the fuzzy MTTF numbers to real values. The defuzzified central (Defvalue) of triangular fuzzy numbers $\tilde{R}=(a,b,c)$ can be determined from (16). The results of the defuzzified values for all the converters are provided in Table I.

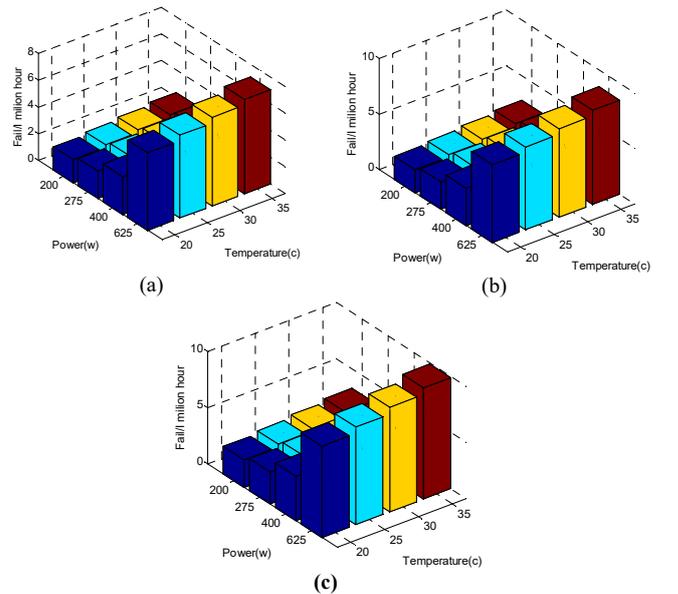

Fig. 5. (a) $ap_{\lambda_i^{(k)}}$, (b) $bp_{\lambda_i^{(k)}}$, and (c) $cp_{\lambda_i^{(k)}}$ values for different power and temperature scenarios (one-phase operation mode).

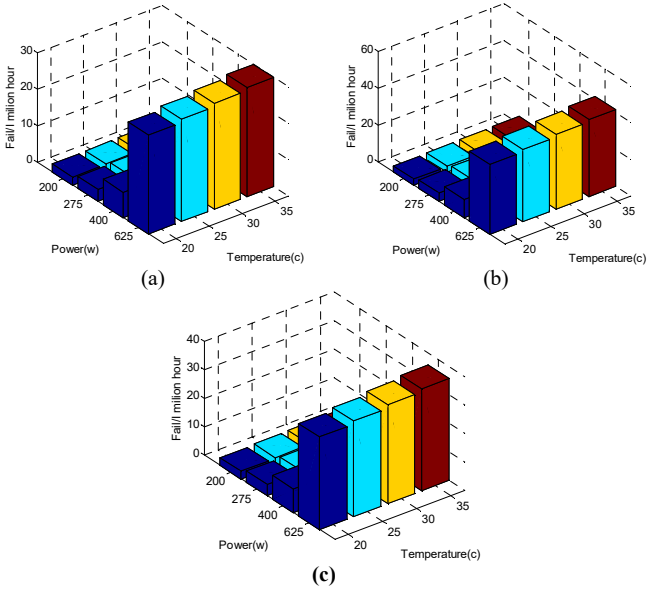

Fig. 6. (a) $as_{\lambda_i(k)}$, (b) $bs_{\lambda_i(k)}$, and (c) $cs_{\lambda_i(k)}$ values for different power and temperature scenarios (one-phase operation mode).

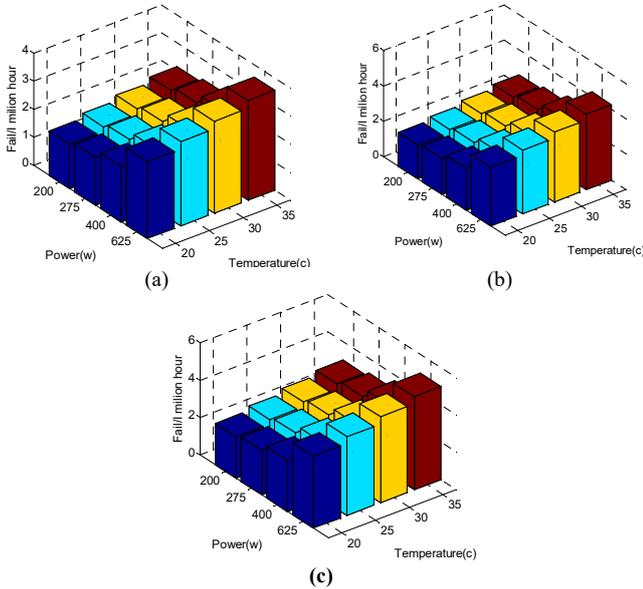

Fig. 7. (a) $ap_{\lambda_i(k)}$, (b) $bp_{\lambda_i(k)}$, and (c) $cp_{\lambda_i(k)}$ values for different power and temperature scenarios (two-phase operation mode).

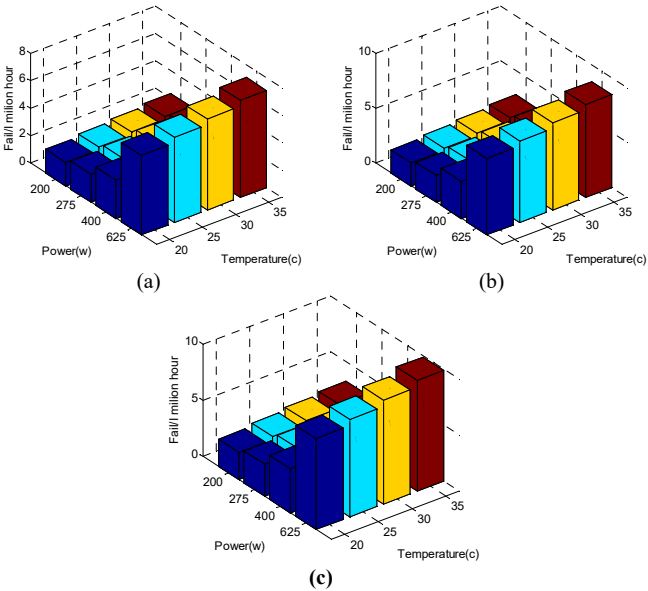

Fig. 8. (a) $as_{\lambda_i(k)}$, (b) $bs_{\lambda_i(k)}$, and (c) $cs_{\lambda_i(k)}$ values for different power and temperature scenarios (two-phase operation mode).

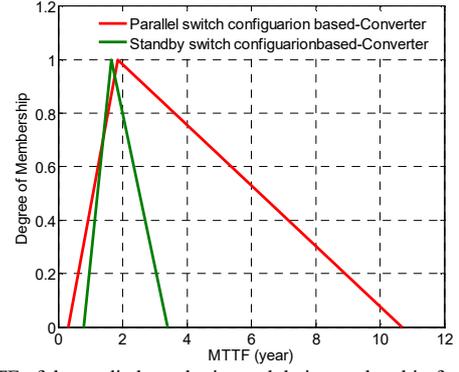

Fig. 9. MTTF of the studied topologies and their membership functions.

Table II. MTTF allocation of different topologies.

| The Studied Topology | Lowest MTTF (year) | Highest MTTF (year) | Defuzzified MTTF (year) |
|---|---|---|---|
| **Parallel Switch Configuration based Converter** | 0.3144 | 10.7 | 4.2871 |
| **Standby Switch Configuration based Converter** | 0.7954 | 3.404 | 1.8536 |

$$Defvalue = \frac{c^2 + bc - a^2 - ab}{3(c-a)} \quad (16)$$

According to Fig. 10 and Table II, the following results can be summarized:

- In the converter with parallel switch configuration, it is likely that the lifetime of the converter is more than 5.7 years. While in the most optimistic case, the lifetimes of the proposed and standby-based converters are 5.7 and 3.4 years, respectively. This means that if the power converter is expected to work for a long time, the parallel switch-based topology should be used.
- In the most pessimistic case, lifetimes of parallel, standby and hybrid switch configuration-based converters are about 0.3144, 0.466, and 0.795 years, respectively. For low mission length, the MTTF of the parallel-switch configuration-based topology is the lowest one and the standby and hybrid switch configuration based-topology are the best topologies to use for low mission length.
- Generally, the lifetime of the parallel-switch configuration-based topology is more than the standby and hybrid ones.

V. CONCLUSION

Fuzzy lifetime Analysis of fault-tolerant the redundant switch configuration-equipped conventional two-phase interleaved converter have been performed in this paper. A converter topology-based Markov chain method is proposed to simplify the reliability and failure Analysis in comparison with the conventional Markov chain-based reliability Analysis that are reported in the literature. Using this method, all possible configurations of the studied converter in different operation modes and healthy modes are determined. MTTF of the studied converter equipped with redundant switch configuration are extracted as functions of power switch failure rates in different operation modes. The illustrated Markov chains consist of the topology of the case study in different operation modes and switch configurations, which give understandable insights to the research on reliability improvements.

## VI. References

[1] F. Blaabjerg, Z. Chen, and S. Kjaer, "Power electronics as efficient interface in dispersed power generation systems," *IEEE Trans. Power Electron.*, vol. 19, no. 5, pp. 1184-1194, Sep. 2004.

[2] H. Khoun Jahan, R. Eskandari, T. Rahimi, R. Shalchi Alishah, L. Ding, K. Bertilsson, M. Sabahi, and F. Blaabjerg, "A limited common-mode current switched-capacitor multilevel inverter topology and its performance and lifetime evaluation in grid-connected photovoltaic applications," *Energies*, vol.14, no.7, pp. 1915-????, Jan. 2021.

[3] A. Abadifard, P. Ghavidel, N. Taherkhani and M. Sabahi, "A Novel Modulation Method to Reduce Leakage Current in Transformerless Z-source PV Inverters," 2020 IEEE 14th Dallas Circuits and Systems Conference (DCAS), 2020, pp. 1-5.

[4] T. Rahimi, H. khoonJahan, F. Blaabjerg, A. Bahman, and S.H. Hosseini, "Fuzzy-logic-based mean time to failure (MTTF) analysis of interleaved dc-dc converters equipped with redundant-switch configuration," *Applied Sciences*, vol. 9, pp. 88, Jan. 2019.

[5] R. Rahimi, et. al. "Filter-Clamped Two-Level Three-Phase Transformerless Grid-Connected Photovoltaic Inverter for Leakage Current Reduction," 2020 IEEE Kansas Power and Energy Conference (KPEC), 2020, pp. 1-6.

[6] D. Zhou, H. Wang, and F. Blaabjerg, "Mission profile-based system-level reliability analysis of dc/dc converters for a backup power application," *IEEE Trans. Power Electron.*, vol. 33, no. 9, pp. 8030-8039, Sep. 2018.

[7] T. Rahimi, L. Ding, R. l. Faraji, M. Kheshti, and J. Pou, "Performance improvement of a three-phase interleaved dc-dc converter without requiring anti-saturation control for post-fault conditions," *IEEE Trans. Power Electron.*, vol. 36, no. 7, pp. 7378–7383, Jul. 2021.

[8] S. V. Dhople, A. Davoudi, A. D. Dominguez-Garci, and P. L. Chapman, "A unified approach to reliability assessment of multiphase dc–dc converters in photovoltaic energy conversion systems," *IEEE Trans. Power Electron.*, vol. 27, no. 2, pp. 739–751, Feb. 2012.

[9] L. Palma and P. N. Enjeti, "A modular fuel cell modular dc-dc converter concept for high performance and enhanced reliability," *IEEE Trans. Power Electron.*, vol. 24, no. 6, pp. 1437-1443, Jun. 2009.

[10] N. Vosoughi Kurdkandi, M. Farhadi, E. Babaei, P. Ghavidel, "Design and analysis of a switched-capacitor DC-DC converter with variable conversion ratio," *International Journal of Circuit Theory and Applications*, Vol. 8, issue 10, pp. 1– 20, 2020.

[11] P. Ghavidel, M. Farhadi, M. Dabbaghjamanesh, A. Jolfaei and M. Sabahi, "Fault Current Limiter Dynamic Voltage Restorer (FCL-DVR) with Reduced Number of Components," in IEEE Journal of Emerging and Selected Topics in Industrial Electronics, 2021.

[12] T. Rahimi, S. H. Hosseini, M. Sabahi, and R. S. Alishah, "EMI consideration of high reliability dc–dc converter: In aerospace based electric transport system charger application," *Prog. Electromagn. Res.M.*, vol. 46, pp. 125-133, MONTH 2016.

[13] M. Farhadi, M. Abapour and B. Mohammadi-Ivatloo, "Reliability analysis of component-level redundant topologies for solid-state fault current limiter", Int. J. Electron., vol. 105, no. 4, pp. 541-558, 2018.

[14] L. Ceccarelli, P. D. Reigosa, F. Iannuzzo, and F. Blaabjerg, " A survey of SiC power MOSFETs short-circuit robustness and failure mode analysis," *Microelectron. Rel.*, vol. 76–77, pp. 272-276, MONTH 2017.

[15] T. Rahimi, S.H. Hosseini, M. Sabahi, G. B. Gharehpetian, and M. Abapour, "Reliability evaluation of a fault-tolerant three-phase interleaved dc-dc boost converter," *Transactions of the Institute of Measurement and Control*, vol. 41, pp. 1278-1289, Mar. 2019.

[16] E. Jamshidpour, P. Poure, and S. Saadate, "Photovoltaic systems reliability improvement by real-time FPGA-based switch failure diagnosis and fault-tolerant dc-dc converter," *IEEE Trans. Ind. Electron.*, vol. 62, pp. 7247-7255, Nov. 2015.

[17] M. Farhadi, M. Abapour, and M. Sabahi, "Failure analysis and reliability evaluation of modulation techniques for neutral point clamped inverters—A usage model approach," *Eng. Failure Anal.*, vol. 71, pp. 90-104, 2016.

[18] T. Rahimi, L. Ding, A. Abadifard, P. Ghavidel, M. Farhadi, R. Faraji, "Unbalanced Currents Effect on the Thermal Characteristic and Reliability of Parallel Connected Power Switches", *Case Studies in Thermal Engineering*, 2021.

[19] S. Peyghami, Z. Wang and F. Blaabjerg, "A guideline for reliability prediction in Power Electronic converters," *IEEE Trans. Power Electron.*, vol. 35, no. 10, pp. 10958-10968, Oct. 2020.

[20] M. Tanrioven, Q. Wu, D. Turner, C. Kocatepe and J. Wang, "A new approach to real-time reliability analysis of transmission system using fuzzy Markov model", *Int. J. Electr. Power Energy Syst.*, vol. 26, no. 10, pp. 821-832, 2004.

[21] X. Bai and S. Asgarpoor, "Fuzzy-based approaches to substation reliability evaluation", *Elect. Power Syst. Res.*, vol. 69, no. 23, pp. 197-204, May 2004

[22] S. J. Aghili and H. Hajian-Hoseinabadi, "Reliability evaluation of repairable systems using various fuzzy-based methods—A substation automation case study", *Int. J. Elect. Power Energy Syst.*, vol. 85, pp. 130-142, 2017

[23] J. Ahn, Y. Noh, S. H. Park, B. I. Choi and D. Chang, "Fuzzy-based failure mode and effect analysis (FMEA) of a hybrid molten carbonate fuel cell (MCFC) and gas turbine system for marine propulsion", *J. Power Sources*, vol. 364, pp. 226-233, Oct. 2017.

[24] S. M. Lavasani, A. Zendegani, and M. Celik, ''An extension to fuzzy fault tree analysis (FFTA) application in petrochemical process industry,'' Process Saf. Environ. Protection, vol. 93, pp. 75–88, Jan. 2015

[25] Komal Naveen Kumar and J. S. Lather, "Reliability analysis of a robotic system using hybridized technique" in Journal of Industrial Engineering International, pp. 1-11, 2017.

[26] Z. G. Li, J. G. Zhou and B. Y. Liu, "System reliability analysis method based on fuzzy probability", *Int. J. Fuzzy Syst.*, vol. 19, no. 6, pp. 1759-1767, 2017.

[27] W. Li, J. Zhou, J. Lu and W. Yan, "Incorporating a combined fuzzy and probabilistic load model in power system reliability assessment", *IEEE Trans. Power Syst.*, vol. 22, no. 3, pp. 1386-1388, Aug. 2007.

[28] D. K. Mohanta, P. K. Sadhu and R. Chakrabarti, "Fuzzy Markov model for determination of fuzzy state probabilities of generating units including the effect of maintenance scheduling", *IEEE Trans. Power Del.*, vol. 20, no. 4, pp. 2117-2124, Nov. 2005.

[29] Y. M. Wang, J. B. Yang, D. L. Xu and K. S. Chin, "On the centroids of fuzzy numbers", *Fuzzy Sets Syst.*, vol. 157, no. 7, pp. 919-926, Apr. 2006.

## VII. Biographies

**Tohid Rahimi** was born in Salmas in Iran, received the BS.c degree from the University of Tabriz, Tabriz, Iran, in 2011, and MS.c degree in electrical engineering in 2013. He received the Ph.D. degree in system and power electronics from the University of Tabriz, Tabriz, Iran, in 2018. He was teacher assistant duration his Ph.D course at University of Tabriz. Also, He was supervisor or advisor several MS.c thesis at the Meraj Higher Education Institute in Iran. He is now a postdoctoral researcher at Shandong University, Jinan, China. He has been served as an active reviewer for different IEEE and other scientific journals. Moreover, he won a competitive grant awarded by China Postdoctoral Science Foundation in 2020. His interests include power electronics, Reliability, EMI, and different fields of power engineering.

**Masoud Farhadi** (S'20) received the B.Sc. degree in electrical engineering with honors and the M.Sc. degree in power engineering (power electronics and systems) with honors from the Department of Electrical Engineering, University of Tabriz, Tabriz, Iran, in 2013 and 2016, respectively. He is currently working toward the Ph.D. degree at the University of Texas at Dallas, Richardson, TX, USA.
His current research interests include analysis and control of power electronic converters, reliability of power electronic systems, wide bandgap semiconductor device's reliability, and renewable energy conversion systems.

**JOSEP POU** (Fellow, IEEE) received the B.S., M.S., and Ph.D. degrees in electrical engineer- ing from the Technical University of Catalonia (UPC)-

Barcelona Tech, in 1989, 1996, and 2002, respectively. He is currently a Professor with Nanyang Technological University (NTU), Singapore, where he is Program Director of Power Electronics with the Energy Research Institute at NTU (ERI@N) and Co-Director of the Rolls-Royce @ NTU Corporate Lab. His research interests include power electronics, renewable energy, energy storage, HVDC systems, and more-electrical aircraft and vessels. He received the 2018 IEEE Bimal Bose Award for Industrial Electronics Applications in Energy Systems. He is also an Associate Editor of the IEEE JOURNAL OF EMERGING AND SELECTED TOPICS IN POWER ELECTRONICS. He was the Co-Editor-in-Chief and an Associate Editor of the IEEE TRANSACTIONS ON INDUSTRIAL ELECTRONICS.

**Ka-Hong Loo** (S'97–M'99) received the B.Eng. (Hons.) in electronic engineering and the Ph.D. degree from the University of Sheffield, U.K., in 1999 and 2002, respectively. Upon completion of his doctoral degree, he won the Japan Society for the Promotion of Science (JSPS) Postdoctoral Fellowship and worked as postdoctoral researcher at Ehime University, Japan, from 2002 to 2004. He joined The Hong Kong Polytechnic University in 2006 where he is now an Associate Professor in the Department of Electronic and Information Engineering. His research interests include high-frequency power conversion, in particular power converters for renewable energy systems. He has been an Associate Editor for the IEEE Transactions on Energy Conversion since 2013 and IEEE Power Engineering Letters since 2015 and contributes regularly as reviewer for various international journals and conferences. He is currently the Chair of the Power Electronics and Control Sub-Committee of the IEEE Technical Committee on Transportation Electrification.